\documentclass[11pt,a4paper]{article}
\usepackage{latexsym}
\usepackage{graphics,graphicx}
\usepackage{amsmath,amsfonts,amssymb,indentfirst,mathrsfs,bm,fancyhdr}
\usepackage[utf8]{inputenc}
\usepackage[T1]{fontenc}
\usepackage{esint}
\usepackage{epsfig}
\usepackage{cite}
\usepackage{hyperref}
\usepackage{color}
\usepackage{float}

\newcommand{\be}{\begin{equation}}
\newcommand{\ee}{\end{equation}}

\def\bea{\begin{eqnarray}}
\def\eea{\end{eqnarray}}

\def\beqx{\begin{displaymath}}
\def\eeqx{\end{displaymath}}

\newcommand{\bmat}{\left(\begin{array}}
\newcommand{\emat}{\end{array}\right)}

\def\i{\iota}

\def\bo{{\raise-.3ex\hbox{\large$\Box$}}}               % D'Alembertian
\def\face{{\raise.2ex\hbox{$\displaystyle \bigodot$}\mskip-2.2mu \llap {$\ddot
        \smile$}}}                                   % happy face
\def\>{\rangle}                                      %right angle
\def\<{\langle}                                      %left angle

\def\leftrightarrowfill{$\mathsurround=0pt \mathord\leftarrow \mkern-6mu
        \cleaders\hbox{$\mkern-2mu \mathord- \mkern-2mu$}\hfill
        \mkern-6mu \mathord\rightarrow$}        % <--> double differential
\def\dvec#1{\vbox{\ialign{##\crcr
        \leftrightarrowfill\crcr\noalign{\kern-1pt\nointerlineskip}
        $\hfil\displaystyle{#1}\hfil$\crcr}}}           % <--> accent
\def\-{\hphantom{-}}

\title{\vskip -0.5in
\begin{minipage}{6.5in}
\begin{flushright}
 {\small IFT UAM/CSIC-14-055}
\end{flushright}
\end{minipage}\\
\vskip 1.0in
Universality of Anomalous Transport for zero temperature Superfluids}

\author{Luis Melgar\footnote{luis.melgar@csic.es}\\ \\
{\it Instituto de F\'{\i}sica Te\'orica UAM/CSIC} \\ {\it Universidad Aut\'onoma de Madrid}\\ {\it Madrid 28049, Spain}}

\begin{document}

\maketitle

\begin{abstract}
We show that generic $U(1)$ superfluids with a $U(1)^3$ anomaly feature universal anomalous transport at low temperature. This universal behavior had been encountered before for a class holographic models by performing explicit computations: anomalous conductivities were found to either vanish or to be $1/3$ of the value they present for ordinary fluids. In this note we argue in favor of
the fact that at zero temperature chiral conductivities are fully determined by anomaly coefficients. We also compute them and show explicitly where their particular value arises from. The proof is based on Ward Identities plus the physical input that at zero temperature all the fluid is in its superfluid component.
\end{abstract}

\newpage

%\tableofcontents
%\addtocontents{toc}{\protect\setcounter{tocdepth}{3}}
\renewcommand{\theequation}{\arabic{section}.\arabic{equation}}
\section{Introduction}
It is been known for some time now that chiral and gravitational anomalies \cite{Bertlmann:1996xk} lead to interesting transport phenomena in systems at finite temperature and density. For ordinary fluids anomalous transport is well-understood in terms of two new basic effects.
One the one hand, there exists the so-called Chiral Magnetic Effect (CME), by which an external magnetic field generates a current parallel to it. Similarly, the Chiral Vortical Effect (CVE) implies that a vortex induces a current parallel to the vorticity. The corresponding 
conductivities are termed Chiral Magnetic and Chiral Vortical Conductivities (CMC and CVC respectively). Both the CMC and CVC have been found to be universal and entirely fixed by the chiral and mixed anomalies of the microscopic theory \cite{Fukushima:2008xe, Kharzeev:2009pj,Son:2009tf, Neiman:2010zi,Newman:2005hd, Erdmenger:2008rm, Banerjee:2008th,Amado:2011zx, Yee:2009vw, Landsteiner:2011cp,Landsteiner:2011iq, Loganayagam:2012pz, Loganayagam:2011mu, Jensen:2012kj}. \\

A very common phenomenon at low temperatures for real systems is to undergo a phase transition to a superfluid phase, in which a global symmetry is broken. General anomalous transport in superfluids has been studied some years ago \cite{Metlitski:2005pr, Newman:2005as, Son:2004tq} and also more recently \cite{Lublinsky:2009wr,  Lin:2011aa, Lin:2011mr, Bhattacharya:2011tra, Bhattacharyya:2012xi, Chapman:2013qpa}. The main conclusion of those works is that
chiral condutivites are now not reduced just to CMC and CVC; two new chiral effects are allowed: The Chiral Electric Effect (CEE)\cite{Neiman:2011mj} induces a current whenever an external electric field is orthogonal to the superfluid velocity and the Chiral Charge Generation Effect (CCGE) implies the presence of a charge density if a magnetic field is aligned with the superfluid velocity\cite{Bhattacharyya:2012xi, JimenezAlba:2014pea}. 
The other main observation is that anomalous transport coefficients are not universal anymore, but they depend on certain thermodynamic functions \cite{Chapman:2013qpa}. However, it has been shown by holographic methods that all of them seem to recover a universal result at zero temperature, at least for a certain class of models. For several models of a $U(1)$ superfluid with a $U(1)^3$ anomaly one finds \cite{Amado:2014mla, JimenezAlba:2014pea}
\begin{align}
\sigma_{CME} (T \rightarrow 0) =&\ \frac{\sigma_{CME}(T>T_c)}{3}\,,\\
\sigma_{CVE} (T \rightarrow 0) =&\ 0\,,\\
\sigma_{CEE} (T \rightarrow 0) =&\ -\frac{\sigma_{CME}(T>T_c)}{3}\,,\\
\sigma_{C^2GE} (T \rightarrow 0)=&\ -\frac{\sigma_{CME}(T>T_c)}{3}\,,
\end{align}
where $T_c$ is the critical temperature and $\sigma_{CME} ,\ \sigma_{CVE},\ \sigma_{CEE}$ and $\sigma_{C^2GE}$ are the CMC, CVC, Chiral Electric Conductivity (CEC) and Chiral Charge Generation Conductivity (CCGC), respectively. \\
It is thus important to clarify whether the anomalous transport coefficients are indeed universal at zero temperature generically (i.e. beyond the holographic approach)  and to understand where their value at $T=0$ comes from. In this note we show that chiral transport coefficients are universal at zero temperature for any $U(1)$ superfluid featuring a $U(1)^3$ anomaly. We are also capable
of recovering the factor of $1/3$ observed in holography. The argument is very general and relies on Ward Identities on the one hand, and the fact that there is no normal component of the fluid at zero temperature (which allows us to constraint the 3-point functions), on the other. \\

In Section \ref{sec:Genst} we present the general strategy. Then we illustrate its implications in the framework of the stationary limit of a 3+1-dimensional chiral superfluid. With this in mind, we comment on the effective action of \cite{Bhattacharyya:2012xi} in Section \ref{sec:effeac}. Then we elaborate on chiral correlators in Section \ref{sec:corrs} to move afterwards to 
the actual computation of zero temperature Chiral Magnetic and Chiral Charge Generation Conductivities (Section \ref{sec:CMCCCGC}), and CVC and $\sigma^{\epsilon}_{C^2GE}$ (Section \ref{sec:CVCCCGCep}). In Section \ref{sec:CEEStreda} we explain why we expect the CEC to equal the CCGC at zero temperature and point out the surprising resemblance with the Streda formula. We will restrict ourselves to $\zeta^2=0$ in most cases. The reason is that, even though we expect the argument to hold
also at finite supervelocity, the calculations involve subtleties related to the way in which we impose the on-shell condition of the Goldstone mode\footnote{A naive derivation from
some of the one-point functions of \cite{Bhattacharyya:2012xi} leads to 3-point functions whose pole-part does not feature the expected symmetries. Moreover, at finite supervelocity we would need to solve the Goldstone phase to second order in the sources in order to extract the form of the general 3-point function. We can take advantage of the fact that the only crucial part of the 3-point function, for the argument presented here, is the one that does not present a pole at zero momentum. Such part is easy to compute at zero $\zeta^2$ and does serve to illustrate the validity of the argument. We will therefore stick just to those kind of terms in general.}. We finish in Section \ref{sec:Conc} with a summary of the present work.
\section{General strategy}
\label{sec:Genst}
We consider a $U(1)$ superfluid with a $U(1)^3$ anomaly. Ward identities for the consistent current read
\begin{align}
\label{WI} \nabla_{\mu} J^{\mu} = -\frac{C}{24} \epsilon^{\mu \nu \rho\sigma} F_{\mu \nu} F_{\rho \sigma} 
\end{align}
where $C$ is the anomaly coefficient. At $\omega=0$, we assume a 2-point function (to first order in $k$) of the form \cite{Jensen:2012jy}
\begin{align}
\label{2point} G^{i, j}_R (\textbf{k})= - i\epsilon^{ijk} k_k  \left(\sigma_{CME} - \frac{C A_0}{3}\right)
\end{align}
Notice that $\sigma_{CME}$ is the Chiral Magnetic Conductivity. On top of that, the 3-point function with an insertion of two spatial components and one temporal component reads
\begin{align}
\label{RC}\nonumber G^{i,j,0}_R (\textbf{k}_1,\textbf{k}_2) = - i \epsilon^{ijk}\Sigma \left((k_1)_k  -(k_2)_k\right)  \\-i \overline{\Sigma}\left(\epsilon^{jkl}\frac{1}{k_1^2} (k_1)^i (k_2)_k (k_1)_l+\epsilon^{ikl}\frac{1}{k_2^2} (k_2)^j (k_1)_k (k_2)_l \right)
\end{align}
where we have used the fact that $\left<...\mathcal{O}^I(\textbf{k}_1)....\mathcal{O}^J(\textbf{k}_2)...\right> = \left<...\mathcal{O}^J(\textbf{k}_2)....\mathcal{O}^I(\textbf{k}_1)...\right>$ for bosonic operators. Moreover, the structure proportional to $\overline{\Sigma}$ arises due to the presence of the Goldstone mode.\\
By means of Ward Identities we can relate (\ref{WI}) to (\ref{RC}) as follows
\begin{align}
i(k_1)_i G^{i,j,0}_R (\textbf{k}_1,\textbf{k}_2) = \epsilon^{jkl}(k_1)_k (k_2)_l \Sigma - \epsilon^{jkl}(k_1)_k (k_2)_l \overline{\Sigma} =\epsilon^{jkl} (k_1)_k (k_2)_l \frac{C}{3}
\end{align}
yielding 
\begin{align}
\boxed{\label{Main}\Sigma - \overline{\Sigma} = \frac{C}{3}}
\end{align}
Remarkable enough, (\ref{Main}) must hold at every point in phase space, no matter if we sit in a broken or unbroken phase. There seems to be a tension between the normal component and the superfluid (overbarred) one. 
\\ We are now in the position of taking the explicit variation of (\ref{2point}) with respect to $A_0 \equiv \mu$ to get
\begin{align}
- i\epsilon^{ijk} (k_1)_k \left(\frac{\partial\sigma_{CME}}{\partial \mu} - \frac{C}{3}\right)= G^{i,j,0}_R (\textbf{k}_1,-\textbf{k}_1) = - i\epsilon^{ijk} (k_1)_k 2 \Sigma\,.
\end{align}
This reasoning leads to 
\begin{align}
\frac{\partial\sigma_{CME}}{\partial \mu} - \frac{C}{3} = 2\Sigma  =  \frac{2 C}{3}+ 2\overline{\Sigma}\,.
\end{align} 
 Notice that in the unbroken phase $\overline{\Sigma} =0$ and we recover the usual result  $\Sigma=\frac{C}{3}$, yielding automatically 
\begin{align}
\frac{\partial\sigma^{\text{unbrok.}}_{CME}}{\partial \mu} = C\,,
\end{align}
as expected. However, in general now we have
\begin{align}
\boxed{\frac{\partial\sigma_{CME}}{\partial \mu} = C + 2\overline{\Sigma}}
\end{align}
In order to further constraint the value of $ \sigma_{CME}$, we need to know $\overline{\Sigma}$, which, in general, will depend on temperature. At zero temperature, however, we could expect it to be a pure number determined by the anomaly. Assuming that at $T=0$ there is no normal component of the fluid and making use of equation (\ref{Main}) it is natural to assume
%\fbox{\parbox{\textwidth}{%
\begin{align}
\label{normal0}  \Sigma (T\rightarrow 0)=0\,,\\
\label{Sigmabar} \overline{\Sigma} (T\rightarrow 0) =  - \frac{C}{3}\,. 
\end{align}
%}}
The above value would mean that at zero temperature all the contribution comes from the superfluid, i.e. $\Sigma\ (T>T_c)=\frac{C}{3}$ turns into $ \overline{\Sigma}\ (T\rightarrow 0)=-\frac{C}{3} $ and $\Sigma\ (T\rightarrow 0)=0$. Substituting (\ref{Sigmabar}) we get automatically
\begin{align}
\frac{\partial\sigma_{CME}}{\partial \mu} (T\rightarrow 0) =  \frac{C}{3}
\end{align}
which is the expected value for the CMC in a superfluid at zero temperature \cite{Amado:2014mla, JimenezAlba:2014pea}.\\

 In subsequent sections we carry out a derivation of 3-point functions in the framework of the stationary effective action for the $U(1)$ superfluid. This allows us to apply the above argument to the remaining superfluid anomalous conductivities in a consistent way.    
\section{Effective action for a 3+1-dimensional superfluid}
\label{sec:effeac}
Contrary to the case of ordinary fluids, the fact that the Goldstone boson $\phi$ is a dynamical massless field forces us to consider an effective action for the Goldstone mode if we want to ensure locality, as opposed to just a
equilibrium partition function in terms of external sources. Such an effective action analysis was undertaken in \cite{Bhattacharyya:2012xi}. Since we are going to use it repeatedly, in this section we review briefly the construction.  
We will be working with a superfluid arising from the spontaneous breaking of a $U(1)$ global symmetry. This implies that the background features a massless phase that corresponds to the Goldstone mode. Different vacua differ by the presence of these Goldstone modes at zero momentum.
The fact that the goldstone mode is massless automatically implies that it has certain impact on the thermodynamics and the hydrodynamics as well, which in turn allows us to define the so-called two-fluid picture, in which two different species of fluids (termed ``normal`` and ``superfluid'' components) coexist even in equilibrium.
After weakly gauging the theory, the system realizes the gauge symmetry
\begin{align}
 \phi \rightarrow \phi +\alpha;  \hspace{1cm} A_{\mu} \rightarrow A_{\mu} - \partial_{\mu} \alpha\,,
\end{align}
which gives rise to a ``London-type'' gauge-invariant source of the form
\begin{align}
 \xi_{\mu} = -\partial_{\mu} \phi + \mathcal{A}_{\mu}\,.
\end{align}
We can see from here that superfluid hydrodynamics will be expressible in terms of $u^{\mu}(x), \xi^{\mu}(x)$ and $T(x)$. It is frequently useful to define the supervelocity as $\zeta^{\mu} = P^{\mu \nu} \xi_{\nu}$, where $P^{\mu \nu}$ is the extrinsic curvature of the induced metric on the hypersurface orthogonal to $u^{\mu}$. This implies $\zeta_i = - \partial_ i \phi + A_i$. We will source the system by putting it in the background of a general stationary metric
\begin{align}
 ds^2 =& e^{-2 \sigma(\vec x)}\left(dt + a_i (\vec x) dx^i\right)^2 + g_{ij}(\vec x) dx^i dx^j\,,\\
\mathcal{A} =& \mathcal{A}_0 (\vec x) dt + \mathcal{A}_i(\vec x) dx^i\,,
\end{align}
with the background values $\mathcal{A}^{\mu}_{(0)} = (\mu_0, \zeta^i_0)$, $a^{(0)}_i = 0, \sigma^{(0)} =0, g^{(0)}_{ij} = \delta_{ij}$. 
We will work with the notation of \cite{Chapman:2013qpa}, i.e.
\begin{align}
\hat T= T_0 e^{-\sigma}\ ,\hspace{0.5cm} \hat \mu = A_0 e^{-\sigma}\ ,\hspace{0.5cm} \hat u^{\mu} = (1,0,0,0)e^{-\sigma}\,.
\end{align}
We will be frequently performing Fourier transformations, that we define as
\begin{align}
\Phi(x) = \int \frac{d^d k}{(2 \pi)^d} \Phi (k) e^{i k_{\mu}x^{\mu}}
\end{align}
Moreover, we define
\begin{align}
\hat \nu \equiv \frac{\hat \mu}{\hat T} = \frac{A_0}{T_0}\ ,\hspace{1cm}
\psi \equiv \frac{\zeta^2}{\hat T^2}\,.
\end{align}
With this ingredients at hand, we can write down the most general \emph{equilibrium} effective action for $\phi$ up to first order in derivatives \cite{Bhattacharyya:2012xi}
\begin{align}
 S=& S_0 + S^{\text{even}}_1 + S^{\text{odd}}\,,\\
S_0= &\int d^3 x \sqrt{g_3} \frac{1}{\hat T} P(\hat T, \hat \mu, \zeta^2)\,,\\
S^{\text{even}} =&\int d^3 x \sqrt{g_3}f\left[ c_1 \left(\zeta\cdot \partial\right) \hat T + c_2 \left( \zeta \cdot \partial \right) \hat \nu + c_3 \left(\zeta \cdot \partial \right)\zeta^2\right]\,,
\end{align}
and 
\begin{align}
S^{\text{odd}}=& S^{\text{odd}}_1+S^{\text{anom}}\,,\\
\label{Sodd} S^{\text{odd}}_1 =&\int d^3 x \sqrt{g_3}\left(g_1 \epsilon^{ijk}\zeta_i\partial_j A_k +T_0 g_2 \epsilon^{ijk}\zeta_i\partial_ j a_k \right)\,,\\
S^{\text{anom}} =& \frac{C}{2} \left(\int \frac{A_0 }{3 T_0} A dA + \frac{A^2_0}{6 T_0} A da\right)\,,
\end{align}
where 
\begin{align}
 g_1 = g_1(\hat T, \hat \nu, \psi); \hspace{1cm}   g_2 = g_2(\hat T, \hat \nu, \psi)
\end{align}
are the thermodynamic functions we previously referred to and we have defined the following integral $\frac{1}{2} \int X dY \equiv \int d^3x \sqrt{g_3} \epsilon^{ijk} X_i \partial _j Y_k$. 
The one-point functions of the current and energy-momentum tensor that we will use read
\begin{align}
 J^i(\vec x) =& \frac{T_0}{\sqrt{g_4}}\frac{\delta W}{\delta A_i}\,,\\
J_0(\vec x) =& -\frac{T_0e^{2 \sigma(\vec x)}}{\sqrt{g_4}}\frac{\delta W}{\delta A_0}\,,\\
T^i_0(\vec x)=& \frac{T_0}{\sqrt{g_4}}\left(\frac{\delta W}{\delta a_i(\vec x)}-A_0(\vec x) \frac{\delta W}{\delta A_i(\vec x)}\right)\,,
\end{align}
being $W = \ln Z$ the generating functional. For instance, one can compute the parity-odd covariant (tilded) contribution to the current as
%\footnote{Henceforth we will assume that $\sigma(\vec x)=0$ and $g_{ij}(\vec x)=\delta_{ij}$ because we will not need to source these fields.} }
\begin{align}
 \delta \tilde{J}_0 =&-e^{\sigma}\left( g_{1 \hat \nu} S_1  + T_0 g_{2,\hat \nu} S_2\right)\,,\\
\label{oddconscurrent}\nonumber \delta \tilde{J}^i =& \hat T \left(2g_1 V^i_6+ T_0 g_2 V^i_7 + \hat T g_{1,\hat T} V^i_1-   \frac{1}{T_0} g_{1,\hat \nu} V^i_2 - g_{1,\psi} V^i_5\right)\\&+ \frac{2}{\hat T}\zeta^i \left(S_1 g_{1,\psi} + T_0 g_{2,\psi} S_2\right) + C e^{-\sigma} \left[2 A_0 V^i_6+ \frac{A^2_0}{2}V^i_7\right]\,,
\end{align} 
where 
\begin{align}
\nonumber S_1 =& \epsilon^{ijk}\zeta_ i \partial_j \zeta_k\,, \hspace{1cm} S_2 = \epsilon^{ijk}\zeta_ i \partial_j a_k\\
V^i_1 =& \epsilon^{ijk}\zeta_ j \partial_k \sigma\,, \hspace{1cm} V^i_2 = \epsilon^{ijk}\zeta_ j \partial_k A_0\,, \hspace{1cm} V^i_5 = \epsilon^{ijk}\zeta_ j \partial_k \psi\,,\\
\nonumber V^i_6 =&\epsilon^{ijk} \partial_ j A_ k\,, \hspace{1cm} V^i_7 = \epsilon^{ijk} \partial_j a_k\,.
\end{align}
Notice that $S_0$ already incorporates the Ward identities. As a consequence the 3-point functions satisfy them trivially and we will not have to impose them in what follows.
\section{Chiral correlators}
\label{sec:corrs}
We will be interested in studying particular two- and three-point functions, associated to anomalous transport. On the one hand, at finite temperature and density
Chiral Magnetic and Chiral Vortical effects exist. Those have associated conductivities that can be computed through the following correlators \cite{Chapman:2013qpa, JimenezAlba:2014pea}
\begin{align}
\label{KuboCME} \sigma_{CME} =& \lim_{k_l \rightarrow 0 } - \frac{i}{2 k_l} \epsilon_{iml}\left< \tilde{J}^i(k) J^m(-k)\right>_{||} = 2 T g_1 +C \mu_0\,,\\
\label{KuboCVE} \sigma_{CVE} =& \sigma^{\epsilon}_{CME} = \lim_{k_l \rightarrow 0 }- \frac{i}{ k_l} \epsilon_{iml} \left< \tilde{J}^i(k) T^m_0 (-k) \right>_{||} = C \mu^2_0 + 4T_0\mu_0 g_1 - 2 T^2_0g_2\,,\\
\label{KuboCMEep} \sigma^{\epsilon}_{CVE} =& \lim_{k_l \rightarrow 0 }- \frac{i}{2 k_l} \epsilon_{iml} \left< T^i_0(k) T^m_0 (-k) \right>_{||} = \frac{1}{3} C \mu^3_0- 2 T^2_0 \mu_0 g_2 + 2 T_0 \mu^2_0g_1 \,,
\end{align}
where the subindex $||$ ($\perp$) means that the correlators must be computed for $\vec \zeta_0 \ ||\ \vec k$ ($\vec \zeta_0 \ \perp \ \vec k$). On the other hand, the presence of the superfluid allows for more chiral effects, such as the Chiral Charge Generation effects, whose correlator read\footnote{We comment on the zero temperature behavior of the Chiral Electric Effect in Section \ref{sec:CEEStreda}.}
%\sigma_{CEE} =& \lim_{\omega \rightarrow 0 } - \frac{i}{\omega} \left< J^y(k) J^x(-k)\right>\,,\\
\begin{align}
\label{KuboCCGE} \sigma_{C^2GE} =& \lim_{k \rightarrow 0 }-\frac{i}{\zeta_{[s}k_{p]}}\epsilon_{spm} \left< J_0(k) J^m(-k)\right>_{\perp} = g_{1,\nu}\,.
\end{align}
Moreover, there exists a related Vortical effect that also induces the presence of a charge density and can be computed in Linear Response Theory using
\begin{align}
\label{KuboCCGEep}  \sigma^{\epsilon}_{C^2GE} =&\lim_{k \rightarrow 0 } - \frac{i}{2 \zeta_{[s} k_{p]}} \left< J_0(k)T^m_0(-k)\right>_{\perp}= T_0 g_{2, \nu}- \mu_0 g_{1,\nu}\,.
\end{align}
Despite the seemingly large collection of anomalous conductivities to study, the general theory of \cite{Bhattacharyya:2012xi} summarized in the previous section reduces the amount of them to the value of two thermodynamic functions, termed $g_1(\mu, T, \zeta^2), g_2(\mu, T, \zeta^2)$. Our aim is to prove that those functions have
universal (i.e. anomaly-constrained) values at $T_0=0$ and to compute such universal value.
\section{The Chiral Magnetic and Chiral Charge Generation Conductivities}
\label{sec:CMCCCGC}
In order to compute $g_1( T_0=0, \nu_0, \psi)$ we need the 3-point function
\begin{align}
\label{3point}\frac{\delta S^{\text{odd}}}{\delta A_l(-k_1) A_m(-k_2) \delta A^0(k_1+k_2)}|_{sources=0}
\end{align}
to first order in momentum. We obtain it directly from taking the necessary variations of the \emph{consistent} charge density $J_0$ (see \cite{Bhattacharyya:2012xi}). This procedure yields
\begin{align}
 \nonumber \left< J^l(k_1) J^m(k_2)J_0 (-(k_1+k_2)) \right>_{(0)} = i \epsilon^{ljm} \left[(k_1)_j- (k_2)_j \right] \left[ g_{1,\nu}+ \frac{C}{3}\right] \\\nonumber + i g_{1,\nu} \left[\epsilon^{ijm}\frac{(k_1)_i(k_2)_j(k_1)^l}{k^2_1} + \epsilon^{ijl} \frac{(k_1)_j (k_2)_i (k_2)^m}{k^2_2} \right]+ 2i g_{1,\nu \zeta^2}\zeta_ i \left[\epsilon^{ijm}\zeta^l(k_2)_j+ \zeta^m \epsilon^{ijl}(k_1)_j\right]\\\nonumber 
-2i \zeta_i \zeta^s g_{1,\nu \zeta^2}\left[\epsilon^{ijm} (k_2)_j (k_1)_ s \frac{(k_1)^l}{k^2_1} + \epsilon^{ijl} (k_1)_j (k_2)_ s \frac{(k_2)^l}{k^2_2}\right] + \mathcal{O}(k^2) \\= i \epsilon^{ljm} \left[(k_1)_j- (k_2)_j \right] \left[ g_{1,\nu}+ \frac{C}{3}\right]+ 2i g_{1,\nu \zeta^2}\zeta_ i \left[\epsilon^{ijm}\zeta^l(k_2)_j+ \zeta^m \epsilon^{ijl}(k_1)_j\right]+ (\text{Pole at } k^2=0) + \mathcal{O}(k^2)
\end{align}
where to the desired order in momentum it is enough to consider the Goldstone solution $\left<\phi \right>^{(1)}_{eq.} = - i \frac{k \cdot \delta A}{k^2}$ (see \cite{Chapman:2013qpa}). Any possible contribution of $\zeta^{(2)}_{i} = -\partial_i \left<\phi\right>^{(2)}_{eq.}$ vanishes after contracting with the epsilon tensor.
From the above equation it follows that $\Sigma$ of Section \ref{sec:Genst} turns out to be
\begin{align}
 \Sigma =  g_{1,\nu}+ \frac{C}{3}\,.
\end{align}
The condition $\Sigma(T=0)=0$ automatically implies
\begin{align}
\label{g1T0} g_1(T \rightarrow 0,\nu, \psi) = - \frac{C}{3}\frac{\mu}{T} + f(\psi)
\end{align}
Notice that for arbitrary supervelocity we also find the extra condition $g_{1,\nu \zeta^2} =0$. \\

Assuming that the CMC behaves smoothly at low temperatures we can restrict $f(\psi) = \text{constant}$. Plugging (\ref{g1T0}) the above value into the Kubo formula (\ref{KuboCME}) we find
\begin{align}
 \label{CMCT0} \lim_{T \rightarrow 0}\  \sigma_{CME} = \frac{C}{3}\mu \,, 
\end{align}
in complete agreement with the holographic computations \cite{Amado:2014mla}. This in particular implies that at zero temperature the Goldstone parity-odd effective action is completely fixed in terms of anomaly coefficients.
Using four-dimensional covariant formulation, we expect it to take the form
\begin{align}
 S^{\text{odd}}_1 \sim - \frac{C}{3} \int d^4 x\   \epsilon^{\mu \nu \rho \lambda} A_{\mu} \zeta_{\nu} \partial_{\rho} A_{\lambda} \,.
\end{align}
On the other hand we find
\begin{align}
 \label{CCGCT0} \lim_{T \rightarrow 0}\  \sigma_{C^2GE} = - \frac{C}{3} \,,
\end{align}
as expected \cite{JimenezAlba:2014pea}.
\section{Chiral Vortical Conductivity and $\sigma^{\epsilon}_{C^2GE}$}
\label{sec:CVCCCGCep}
From the previous section it is clear that in practice we need to identify the part that lacks a pole at zero momentum in the three-point function and then impose that
such contribution vanishes at zero temperature. To find the zero temperature value of $g_2$ we need to analyze the following 3-point function
\begin{align}
\nonumber \left<T^m_0(k_2) J^l(k_1) J^0(-(k_1+k_2))\right>_{\zeta^2=0} = - i(k_2)_j T \epsilon^{lmj}\left[g_{2,\nu} + \frac{C A_0}{3T}\right]+(\text{Pole at } k^2=0) + \mathcal{O}(k^2)\,,
\end{align}
where we have assumed that the background supervelocity is zero in this case, for simplicity, and substituted (\ref{g1T0}). The same reasoning as before thus implies the condition 
\begin{align}
 g_2 (T \rightarrow 0, \nu, \psi)= -\frac{C}{6} \frac{\mu^2}{T^2} + \tilde{f}(\psi)\,.  
\end{align}
The function $\tilde{f}$ cannot be further constrained without performing an analysis of the 3-point function at finite supervelocity. Equipped with the values of $g_1$ and $g_2$ at zero temperature, we can compute the CVC in this limit. This yields 
\begin{align}
 \label{CVCT0} \lim_{T \rightarrow \ 0}\  \sigma_{CVE} =  \lim_{T \rightarrow 0}\ -2 T^2 \tilde{f}(\psi)  \,,
\end{align}
which vanishes at zero supervelocity, matching the holographic result of \cite{Amado:2014mla}. Moreover, equation (\ref{KuboCCGEep}) for $\sigma^{\epsilon}_{C^2GE}$ leads to
\begin{align}
  \label{CCGEepT0} \lim_{T \rightarrow \ 0}\ \sigma^{\epsilon}_{C^2GE} = 0 \,.
\end{align}
\section{The Chiral Electric Conductivity and the Streda formula}
\label{sec:CEEStreda}
The effective action for the Goldstone field has the disadvantage of being stationary. This precludes the computation of the Chiral Electric Effect\footnote{It would be interesting to investigate whether there exists also a vortical-type effect related to the CEE.}, whose associated Kubo formula is (in components)
\begin{align}
\label{KuboCEE} \sigma_{CEE} = \lim_{\omega \rightarrow 0 } - \frac{i}{\zeta_z \omega} \left< J^y(k) J^x(-k)\right>\,.
\end{align}
However, as commented in \cite{JimenezAlba:2014pea}, gauge invariance of the electric field $E_i = \partial_0 A_i -\partial_i A_0$ can come into help for one can use a Kubo formula which is identical to (\ref{KuboCCGE}), instead of (\ref{KuboCEE}). The equivalence between (\ref{KuboCCGE}) and (\ref{KuboCEE}) seems to hold with good approximation at zero temperature \cite{JimenezAlba:2014pea}.
Thus we conclude that the expected value for the CEC at low temperature is
\begin{align}
 \label{valuCEE} \sigma_{CEE} (T= 0) =  \sigma_{C^2GE} (T= 0) = - \frac{C}{3}\,.
\end{align}
Remarkably, this has a striking resemblance with the Streda formula used in Hall-type systems. If $\sigma_H$ is the Hall conductivity, the Streda formula reads
\begin{align}
 \label{Streda} \sigma_H = \left(\frac{\partial B}{\partial \rho}\right)^{-1}\,,
\end{align}
being $B$ the magnetic field and $\rho$ the charge density. The CEE is a Hall-type effect in which the Hall conductivity is proportional to the superfluid velocity, i.e. $j^x_{CEE} \propto \zeta_z E_y $. Furthermore, notice that the right hand side of equation (\ref{Streda}) corresponds to $\sigma_{C^2GC}$. So applying Streda formula
(appropriately generalized to superfluids) it follows that $\sigma_{CEE} = \sigma_{C^2GE}$. \\
We can gain insight into the relation between the Streda formula and the CCGE by the following simple argument. Consider a finite sample of a chiral superfluid with supervelocity $\zeta_z$ and an external electric field $E_y$. Then the existence of the anomaly would imply the presence of a current $j^{CEE}_x = \sigma_{CEE}\ \zeta_z E_y$. Let us neglect 
any other possible contribution to the current for simplicity. On top of this setup we induce now a magnetic field $B_z$. The Hall Effect will then take place\footnote{Or, to be more precise, a London Hall Effect\cite{Greiter:1989qb}.}, inducing a charge excess on the boundary of the sample given by 
\begin{align}
 \rho = \frac{B_z}{E_y} j^{CEE}_x = \sigma_{CEE}\ \zeta_z B_z\,.
\end{align}
 The above picture allows us to regard the CCGE just as a consequence of the combination of the Chiral Electric Effect and the Hall Effect, and makes it explicit that $\sigma_{CEE} = \sigma_{C^2GE}$. It could even be possible that there exist some similar effect related to the Thermal Hall Effect. \\

It would be interesting to prove formula (\ref{valuCEE}) by constructing the full (i.e. non-stationary) effective action.
\section{Conclusions and further analysis}
\label{sec:Conc}
We have shown that any $U(1)$ chiral superfluid features universal (fully anomaly-constrained) anomalous transport coefficients at zero temperature. The proof is based on two assumptions
\begin{enumerate}
\item (Anomalous) Ward Identities hold.
\item At zero temperature there is not normal component of the fluid. 
\end{enumerate}
At the practical level such universal value (see equations (\ref{CMCT0}),(\ref{CCGCT0}),(\ref{CVCT0}),(\ref{CCGEepT0}) and (\ref{valuCEE})) is a consequence of the fact that 
\begin{align}
 g_1 (T=0, \nu,\psi) = -\frac{C}{3}\nu + f(\psi)\ , \hspace{1cm}  g_2 (T=0, \nu,\psi) = -\frac{C}{6}\nu^2+ \tilde{f}(\psi)
\end{align}
Functions $f(\psi)$ and $\tilde{f}(\psi)$ are not determined by the construction because almost every calculation has been performed at zero superfluid velocity. However, we have been able to constraint $f(\psi)= \text{constant}$ by assuming regularity of the CMC as we approach the zero temperature limit. In principle, we expect both $f(\psi)$ and $\tilde{f}(\psi)$ to be
fixed by the procedure presented here once we consider the 3-point functions $\left<J^l(k_1)J^m(k_2)J_0(-k_1-k_2)\right>$, $\left<T^l_0(k_1)J^m(k_2)J_0(-k_1-k_2)\right>$ and $\left<T^l_0(k_1)T^m_0(k_2)J_0(-k_1-k_2)\right>$ at arbitrary superfluid velocity\footnote{We also expect that the tensor structure that arises associated to the supervelocity is independent of that related to the zero supervelocity part, as in Section \ref{sec:CMCCCGC}.}. This much more complicated analysis is beyond the scope of this note. \\ 

Finally, let us point out that similar freedom as the one seen in (\ref{Sodd}) has been found when considering effective theories with dynamical massless modes \cite{Jensen:2013vta}. It could be interesting to investigate whether the strategy presented here
could serve to fix zero-temperature chiral transport coefficient in that case as well. Of course assumption 2 above looses its meaning, but it may still be possible to justify the validity of (\ref{normal0}).
%\newpage
\section*{Acknowledgments}
I am indebted to C. Hoyos for his help and support. I also thank S. Chapman for help with the calculations and K. Jensen, K. Landsteiner, I. Amado, A. Jimenez-Alba and D.T. Son for fruitful discussions and comments on the draft. I would also like to thank the organizers of the ``Cargese Summer School 2014'' and the people from KITPC Beijing for their hospitality during the time 
at which this work was finished. L. M. is supported by Plan Nacional de Altas Energ\'\i as FPA 2009-07890, Consolider Ingenio 2010 CPAN CSD200-00042 and
HEP-HACOS S2009/ESP-2473. L.M. has been supported by FPI-fellowship BES-2010-041571.

%We would like to thank K.Landsteiner especially for illuminating discussions and constant support. We also want to %thank C.Hoyos and I.Amado for useful comments on the draft.  A. J. and L. M. are supported by Plan Nacional de Altas Energ\'\i as FPA
%2009-07890, Consolider Ingenio 2010 CPAN CSD200-00042 and
%HEP-HACOS S2009/ESP-2473. L.M. has been supported by FPI-fellowship
%BES-2010-041571. A. J. has been supported by FPU fellowship AP2010-5686. L.M. thanks Susana Hernández for tremedously joyful discussions.

\newpage

\addcontentsline{toc}{section}{References}

\nocite{*}

\bibliographystyle{jhepcap}
\bibliography{Chirsup}

\end{document}